\documentclass[twocolumn,aps,floatfix,prl,showpacs]{revtex4}
\usepackage{amssymb}

\usepackage{amsmath}
\usepackage{graphicx}

\begin{document}

\title{Verification of stable operation of rapid single flux quantum devices 
with selective dissipation}
\author{J. Hassel, L. Gr\"onberg, and P. Helist\"o}
\affiliation{VTT, P.O. Box 1000, 02044 VTT, Finland}


\begin{abstract}
It has been suggested that Rapid Single Flux Quantum (RSFQ) devices could be
used as the classical interface of superconducting qubit systems. One
problem is that the interface acts as a dissipative environment
for a qubit. 
Recently ways to modify the RSFQ damping to reduce the dissipation
have been introduced. One of the
solutions is to damp the Josephson junctions by a frequency-dependent linear
circuit instead of the plain resistor. The approach has previously been
experimentally tested with a simple SFQ comparator. In this paper we perform
experiments with a full RSFQ circuit, and thus conclude that in terms of
stable operation the approach is applicable for scalable RSFQ circuits.
Realisation and optimisation issues are also discussed.
\end{abstract}

\maketitle

\affiliation{VTT, P.O. Box 1000, 02044 VTT, Finland}

\bigskip

\section{Introduction}

There is an ongoing effort to integrate Rapid Single Flux Quantum\ (RSFQ)
technology \cite{lik1} as the classical interface of qubit systems \cite{sem1,ave1,fed1}%
. Despite the potential compatibility of the fabrication technology and the
operating environment, it has been found that certain issues need to be
resolved before functional RSFQ/Qubit systems are feasible. The basic
question is to develop devices with sufficient functionality, which also
preserve quantum coherence. The generic issues that have been addressed from
the RSFQ side include fabrication \ \cite{cas1, gro1} and design \cite{sav1,int1} solutions
to decrease self-heating \cite{sav2} as well as the control of the
level of dissipation. The latter is important since the classical interface
acts as a dissipative environment for a qubit. The level of dissipation
experienced by the qubit can be reduced by choosing a low enough level of
coupling between the RSFQ circuit and the qubits. This may degrade the
functionality by limiting the signal levels seen by the qubit or by
degrading the readout resolution. There are realisations of particular RSFQ
components, which by design are less dissipative than the conventional ones 
\cite{ave1,fed1,wul1}. Furthermore, unconventional damping schemes have been proposed
which enable the design of generic RSFQ circuits with reduced dissipation.
One such scheme is based on nonlinear shunts \cite{zor1}. Another scheme is
based on linear frequency dependent shunts, for which the damping resistor
has been high-pass filtered by an appropriate circuit \cite{has1}. In the
simplest form the filtering circuit consists of a capacitor $C_{s}$ in
series with the shunt resistor $R_{s}$. The $R_{s}C_{s}$ cutoff is to be
chosen sufficiently below the plasma frequency $\omega_{p}$ of the Josephson
junction (JJ). The principle of $RC$ damping is illustrated in Fig. 1
together with a microscope photograph of a realisation using a Nb trilayer
process \cite{gro1}.

\begin{figure}[tbp]
\includegraphics[width=8cm]{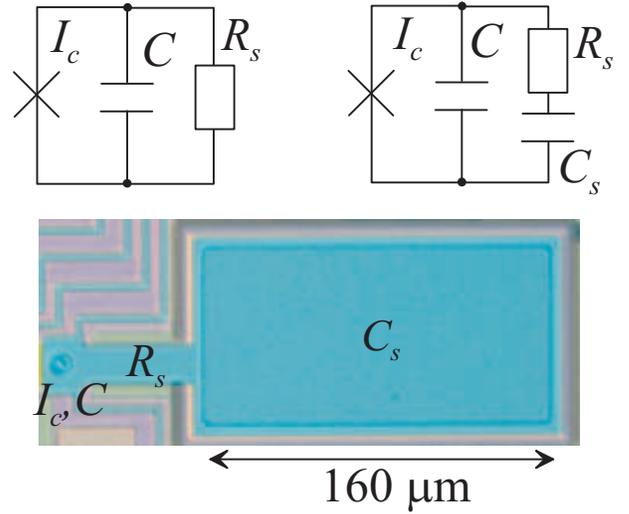}
\caption{Schematics of (a) The conventional and (b) $RC$ damped Josephson
junction. (c) A microscope photograph of an $RC$ shunted Josephson junction
realised by a Nb trilayer process.}
\end{figure}

To our knowledge, the experimental verification of the unconventional
damping schemes is up to date limited to single JJs \cite
{zor1,lot1} or SFQ comparators \cite{has1}, though the functionality of
complete RSFQ\ circuits has been verified by network simulations \cite{has1,
zor2}. In this paper we verify experimentally that complete $RC$ shunted
RSFQ\ devices function in a stable manner.

\section{The Device}

The circuit diagram and a microscope photograph of the device under study
are shown in Fig. 2. The device is a Toggle flip-flop (TFF) driven by a
DC/SFQ converter through a short section of Josephson Transmission Line
(JTL). Apart from the damping arrangement the design of the circuit elements 
has been adopted from \cite{lik1} and \cite{pol1}. The dynamical sequence of the
device is described as follows. As the input current $I_{in}$ of the DC/SFQ
converter is ramped up, at a certain treshold value the phase of junction J3
rotates by 2$\pi$ (J1 flips), which causes a flux quantum to propagate
through the JTL to the TFF. The side-effect is that a persistent current
flowing in loop LP1 changes causing junctions J1 and J2 to flip instead of
J3 during the ramp-down. The ramp-down thus restores the original persistent
current configuration in LP1, but has no other effect completing the DC/SFQ
cycle. As a flux quantum enters the TFF either junctions in pair J8 and J10,
or pair J9 and J11 flip depending on the value of the persistent currents in
the TFF (loops LP2 and LP3). These events also change the value of the
persistent current in LP2 and LP3 in such a way that subsequent events
alternate the persistent current between two values corresponding to zero
and one flux quanta through the loops. The above description thus confirms
that the device utilises all the basic elements of RSFQ dynamics, namely the
selection process as well as the propagation and the storage of flux quanta.
The resulting simulated time-trace of the flux in LP2 is shown in Fig. 3(a).

\begin{figure}[tbp]
\includegraphics[width=8cm]{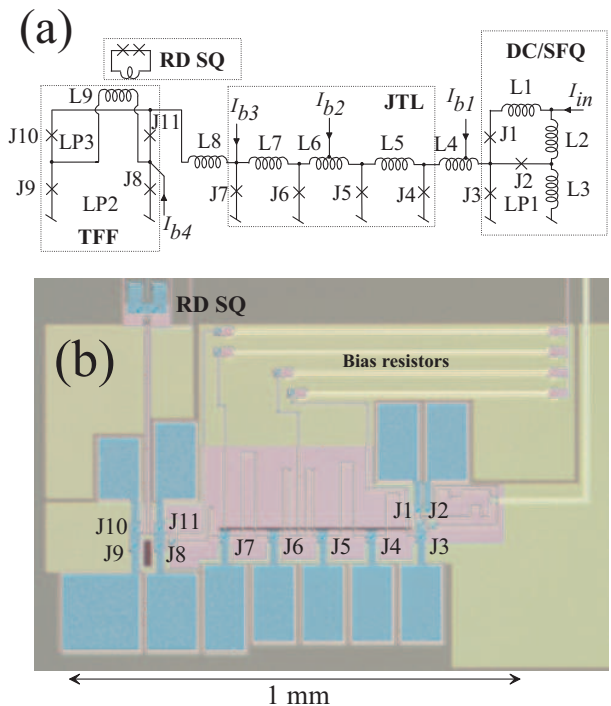}
\caption{The device used for verifying the stable operation of $RC$ shunted
RSFQ. (a) The equivalent circuit. The shunting circuits are not shown. The
critical currents of junctions JN are $I_{c1}$...$I_{c6}$ = $I_{c0}$, $%
I_{c7} = 1.06I_{c0}$, $I_{c8} = 1.12I_{c0}$, $I_{c9}= 1.42I_{c0}$, $I_{c10}=
0.89I_{c0}$ and $I_{c11}= I_{c0}$. Here $I_{c0}\approx 12$ $\protect\mu$A.
The JJ capacitances are $C_n \approx 0.3$ pF/$\protect\mu$A$\times I_{cn}$.
The shunt resistances and capacitances are chosen so that $\protect\beta_c
\approx 0.4$ and $\protect\gamma \approx 0.1$. Inductances are $L_1 = 0.35
L_0$, $L_2 = 0.33 L_0$, $L_3 = 0.6 L_0$, $L_4 = 0.4L_0$, $L_5 = 0.6L_0$, $%
L_6 = L_0$, $L_7 = 0.65L_0$, $L_8 = 0.35 L_0$, and $L_9 = 1.33L_0$. Here $%
L_0\approx$ 94 pH. The bias current ratios are determined by on-chip bias
resistors as $I_{b2}/I_{b1} = 0.95$, $I_{b3}/I_{b1} = 0.77$, $%
I_{b4}/I_{b1} = 0.77$. (b) A microscope photograph of the device.}
\end{figure}

The loop of the DC/SFQ converter is coupled to the loop of a readout dc
SQUID (RD SQ) by a weak inductive coupling used to sense the state of the
TFF. By replacing the RD SQ by a flux qubit, this type of an arrangement
could be utilised for qubit manipulation. Here we are using the TFF - RD SQ
arrangement as a SFQ/DC converter to be able to verify the proper operation
of the circuit.

The device is fabricated by a Nb trilayer process, the VTT\ RSFQubit process 
\cite{gro1} which fixes the critical current density $J_{c}=$ 30 A/cm$^{2}$
for the JJs. The junction size is set to about 7 $\mu$m varying slightly
depending on the design of the elements (see caption of Fig. 2 for details).
All the junctions in the RSFQ circuit are damped with $RC$ shunts. The
hysteresis parameter for the JJs is $\beta_{c}=2\pi I_{c}R_{s}^{2}C/\Phi_{0}$
$\approx0.4$, and the ratio of $R_{s}C_{s}$ cutoff and the JJ plasma
frequency to about $\gamma=1/R_{s}C_{s}\omega_{p}\approx0.1$ \cite{has1}.
This leads to typical physical values for the JJ and the shunt parameters values 
$I_{c}\approx12$ $\mu$A, $C\approx1.9$ pA, $R\approx2.4$ $\Omega$ and $%
C_{s}\approx35$ pF. The inductance values are of the order of 100 pH.

\section{The Experiment}

The experiments were performed at liquid He by a conventional cryoprobe. The
SFQ input signal and the bias currents (the SFQ bias and the RD SQ
current and flux bias) were low-pass filtered by room temperature $RC$ filters, and
fed to the 4.2 K stage through twisted pairs of constantan wire. The RD SQ
voltage output was amplified\ by a commercial room temperature
preamplifier. An experimental plot is shown in Fig. 3(b). The similarity in
comparison with the simulated data of Fig. 3(a) confirms that the device
works in the desired mode described above. The difference in the time scales
is irrelevant, since between the SFQ events the device is in the quiescent
state. The shorter time scale is chosen in the simulation to minimise the
computational effort. The only requirement is that the frequency of the DC/SFQ
input signal $I_{in}$ is much smaller than the inverse of the time scales of
the SFQ events (here of order 100 ps). The only discrepancy between the
simulation and the experiment is that the treshold current is slightly
smaller (about 30 $\mu$A) in the experiment as compared to the simulation
(about 50 $\mu$A). This may be partially explained by the difference in the
designed and realised inductance values, but more likely the cause is
partial flux trapping. The treshold current for a given device varied
somewhat between different cooldowns, which supports this hypothesis. The
treshold variation was also reproduced in simulations by applying flux to
the loops of the DC/SFQ converter. The noise at the output voltage $V_{RD}$
is well explained by the preamplifier voltage noise.

\begin{figure}[tbp]
\includegraphics[width=8cm]{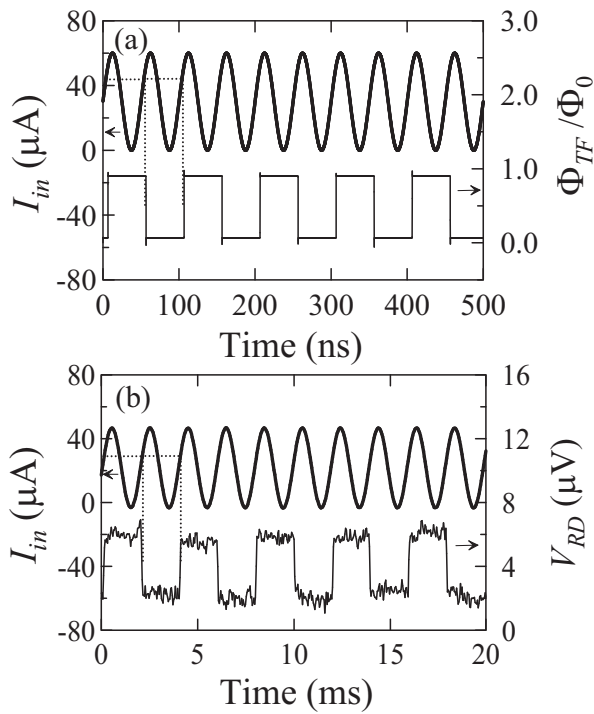}
\par
.
\caption{(a)\ Simulated and (b) measured characteristics of the device. The
thick line (left axis) is the input current of the DC/SFQ converter $I_{in}$%
. The thin line is the flux $\Phi_{TF}$ through loop LP2 of the TFF in (a)
and the output voltage $V_{RD}$ of the readout SQUID in (b) which is
proportional to $\Phi_{TF}$. The bias point is here $I_{b1}\approx 7$ $%
\protect\mu A$}
\end{figure}

\section{Discussion}

Since our main aim is to verify here that the high-frequency damping alone
suffices to stablise the devices, we briefly discuss the excess
damping mechanisms present in our system. The current bias of the device is
based on on-chip bias resistors of $R_{b} \gtrsim  $40 $\Omega $ performing
the required voltage-to-current conversion. The voltage is fed over an
off-chip resistor (0.5 $\Omega $) located near the chip at 4 K. Thus the
bias resistor and the off-chip resistor form a loop of about 40 $\Omega $ in
series with a bonding wire inductance of order $1$ nH, i.e. the excess
damping ranges roughly from DC to 5-10 GHz. As the worst-case the hysteresis
parameter from bias resistors corresponds to $\beta _{c}\sim 100$, which is
not enough to stabilise the system. The RD SQ, which is $R$-shunted,
potentially causes some excess dissipation as well, which couples,
however, to the RSFQ circuit only at high-frequencies ($\gtrsim$10 GHz 
in our circuit). To ensure that 
the excess dissipation mechanisms do not stabilise our system, 
we removed the $RC$ shunts and added the damping from the RD SQ and the
bias resistors in the simulation. The result was that no stable
operating point was found. It thus proved that the stabilisation of our
system is in practise completely due to the $RC$ shunts.

One issue in $RC$ damped circuits is the parasitic resonance
of the shunt capacitor. A sufficient criterion for avoiding this is that the
capacitor dimension should be at maximum $\lambda _{p}/8$, where $\lambda
_{p}$ is the wavelength in the capacitor dielectric at the plasma frequency.
This is given as $\lambda _{p}=2\pi c/\omega _{p}\sqrt{\varepsilon
_{r}\left( 1+2\lambda _{L}/d\right) }$, where $c$ is the speed of light, $%
\varepsilon _{r}$ is the dielectric constant, $\lambda _{L}$ is the London
penetration depth of the electrodes, and $d$ is the insulator thickness. In
our case $\omega _{p}/2\pi =\sqrt{\left( 1/2\pi \Phi _{0}\right) \left(
I_{c}/C\right) }\approx 22$ GHz, $\varepsilon _{r}\approx 45$ (Nb$_{2}$O$_{5}
$ dielectric), $\lambda _{L}\approx $ 85 nm (Nb electrodes) and $d\approx $%
140 nm. It follows in our case that $\lambda _{p}/8\approx 170$ $\mu $m,
which is also the maximum dimension we have used in the capacitors. However,
in our geometry the lowest frequency resonance that could be excited is a $%
\lambda /2$ resonance corresponding to the long edge of the capacitor, so
elements somewhat larger than this could be safely realised. In more general
terms, parasitic resonances limit the maximum value of realizable
capacitance. Since capacitance of a square with side $w$ is $%
C_{s}=\varepsilon _{r}\varepsilon _{0}w^{2}/d$, it follows $C_{s}\lesssim
(\pi c)^{2}\varepsilon _{0}/16\omega _{p}^{2}d\left( 1+2\lambda
_{L}/d\right) $ from the requirement $w\lesssim \lambda _{p}/8$. In other
words, realizability requires that

\begin{equation}
\gamma \gtrsim \frac{16\omega _{p}d\left( 1+2\lambda _{L}/d\right) }{\pi
^{2}R_{s}c^{2}\varepsilon _{0}}=\frac{32\sqrt{2}d\left( 1+2\lambda
_{L}/d\right) }{\pi \Phi _{0}\varepsilon _{0}c^{2}}I_{c},  \label{gmin}
\end{equation}
where we have used the definition of $\gamma $. In the last form we have
also used the definitions of $\omega _{p}$ and $\beta _{c}$, and furthermore
set $\beta _{c}=1/2$, which is \ a typical value. Realizability thus gives
the minimum $\gamma $.

From Eq. (\ref{gmin}) we see that the minimum $\gamma $ depends only $%
\lambda _{L},$ $d$ and $I_{c}$. The maximum is determined from the stability
requirement, which typically leads to $\gamma \lesssim 0.3$ \cite{has1}. The
thickness $d$ can be varied to some extent within fabrication tolerances.
The dependence on $I_{c}$ is, however, more important in terms of the
optimsation of the RSFQ/qubit systems. The realizability and stability
together set an upper limit for $I_{c}$. Low $I_{c}$ is favorable also in
terms of minimal self-heating of RSFQ components \cite{sav1,int1,sav2}. On the
other hand, it was previously analysed that in case of a qubit inductively
coupled to an RSFQ circuit, the minimisation of the dissipation favors an
RSFQ\ circuit with large $J_{c}$ \cite{has1}. Therefore it appears that at
least in this coupling scheme the optimum solution would be a large-$J_{c}$
process with small area junctions, e.g. sub-$\mu$m Nb junctions \cite{pat1}. 
However, there may be other ways
around this as well. It may be possible to relax the realizability restrictions
by different implementations of the damping circuit. It also depends very
much on the particular design whether the self-heating is a problem. For
example, the device measured in this paper spends most of its time in the
quiescent state, whence it should not heat very much above the bath
temperature.

\section{Conclusion}

In conclusion we have successfully verified the stability of a
RSFQ device based on selective
damping realised by $RC$ shunts. The device under study utilises all aspects
of RSFQ dynamics. Our measurement scheme also gives direct evidence on SFQ
events instead of the earlier measurement based on the statistical
properties of a balanced comparator \cite{has1}. We therefore conclude
that it is now experimentally verified that generic RSFQ devices can be
realised by this damping scheme.

\section{Acknowledgment}
The authors wish to thank H. Sepp\"a, M. Kiviranta, and A. Kidiyarova-Shevchenko
for useful discussions. The project was supported by EU through FP6 project
RSFQubit and by the Academy of Finland.

\end{document}